# The relaxation time of OH bond for hydrogen impurity in LiNbO₃

**Pradipta Giri[1], Arindom Biswas[2] and Mrinal Kanti Mandal[1*]**

[1]Department of Physics, National Institute of Technology, Durgapur 713209, India
[2]Department of Electronics and Communication Engineering, Asansol Engineering College, Asansol 713305, India
[*]E-mail: mrinalkanti.mandal@phy.nitdgp.ac.in

**Abstract:** The one dimensional model for the dynamic of hydrogen in lithium niobate is explained by adopting Morse potential. The diffused hydrogen substitutes Lithium and it makes bonding with one oxygen atom of a facet of oxygen-triangle. The bonds will be stretched to set up anharmonic vibration. The damped anharmonic oscillation is derived to explain the dynamics of hydrogen as an impurity. The thermal fluctuation is studied by Fokker Planck equation has an important role to determine the diffusion constant for substitutional hydrogen. The hydrogen diffusion constant and relaxation time are calculated to support the proposed theory and existing experimental results. The concentration of substitutional hydrogens is studied with the help of Boltzmann distribution.

## 1. Introduction

The ferroelectrics such as lithium niobate (LiNbO₃) and lithium tantalite (LiTaO₃) are promising materials in the field of applied physics. These have important applications [1-2] in nonlinear optics, such as electro-optics, second harmonic generation [3-4] and nonvolatile memory devices [5]. Hydrogen is a very common impurity in oxides, generally associated with host $O^{2-}$ ions forming the $OH^-$ molecules. The formation of $OH^-$ group was explained by decay of LiNb₃O₈ during annealing [6] in humid atmosphere at temperature between $400°C - 700°C$. The $(Li_{1-x}H_x)NbO_3$ molecules [7] are produced due to the diffusion of hydrogen in LiNbO₃ is used in fabrication of the optical wave guides for the use in integrated optics [8]. One of the most promising techniques for the fabrication of such wave guides is annealed proton exchange. The hydrogen incorporation increases the resistance to the optical damage [9] and the extraordinary refractive index [10] of LiNbO₃ crystal. The extraordinary refractive index of the birefringe of LiNbO₃ will be increased that can be studied in illumination with a high intensity laser beam. The hydrogen as an impurity in LiNbO₃ is used for fixing the photorefractive effect employed for hologram storage [11-12].

The presence of diffused hydrogen in LiNbO₃ can be monitored using infrared (IR) absorption [13] of $OH^-$ ions which peaks at 3482 cm⁻¹. The vibrational modes of $OH^-$ ions can be interpreted in the framework of an anharmonic oscillator model. The sufficiently broad IR band is observed in LiNbO₃ in presence of hydrogen as an impurity [14]. The broadness of IR spectra and its peak position in LiNbO₃ remarkably depend on the $[Li]/[Nb]$ ratio [15]. With the decrease of Lithium vacancies the position of peak is shifted and band is narrowed due to disappearance of intrinsic defect. This defect is explained that Li atoms have been replaced with hydrogen atoms. The high hydrogen concentration introduced into the host lattice may induce large crystal distortions.

Ferroelectric materials such as $LiNbO_3$ and $LiTaO_3$ show a nonlinear hysteresis in which coercive field ($E_c$) decreases with the smooth increase of Li concentration (mole %) [16]. Again both $E_c$ and internal field ($E_{int}$) decrease with the increase of temperature. The optical observation in $LiNbO_3$ is pronounced that sideways domain wall movement is attributed to pinning of domain walls defects [17]. One dimensional array of a ferroelectric slab of the domain with uniform polarization function of space and time is explained by Klein Gordon (K-G) equation [18] which is used to explain the motion of domain walls. The nonlinear Schroedinger equation (NLSE) results in perturbation analysis of K-G equation. The bright soliton, resides in the vicinity of the defect and also at the domain walls, is the solution of the NLSE [19]. The K-G equation is used to determine the refractive index which is enhanced at the presence of dark soliton [20]. The vapor transport equilibration (VTE) technique demands that OH bonds have responded to ferroelectric switching in near stoichiometric Lithium niobate [21].

Diffusion of hydrogen as an impurity in $LiNbO_3$ is two types. Some diffused hydrogen atoms are employed to substitute Lithium and able to form bonding with oxygen atom of facet of oxygen triangle. These diffused hydrogen atoms are called substitutional hydrogen. Another type is interstitial hydrogen atoms which are small size and commonly observed in metal, semiconductor. Interstitial diffusion coefficient is larger than substitutional diffusion coefficient. Probability of substitutional hydrogen among the diffused hydrogen atoms in $LiNbO_3$ changes with temperature. The Morse potential is appropriate to describe the fundamental energies of transition and higher vibrational transition of OH molecules. So Morse potential is used to study the damped anharmonic oscillation of OH bond in $LiNbO_3$. The diffusion coefficient of substitutional hydrogen is determined from non-dimensional Fokker Planck equation. The damping constant as well as relaxation time is yielded from diffusion coefficient. Solution of Fokker Planck equation is Boltzmann distribution that predicts roughly percentage of substitutional hydrogen in $LiNbO_3$ at different temperature. A part of the diffused hydrogen atoms are responsible to substitute the lithium and this substitutional hydrogen atoms are able to form OH bonds. The formation of OH bonds are considered first in ground state then absorb energy and get excited to higher energy level. So absorption spectra are speculated in $LiNbO_3$. The stretching mode frequency of vibration of OH ions in $LiNbO_3$ are in the range between 3200 $cm^{-1}$ and 3700 $cm^{-1}$.

## 2. The theory

In oxide crystal, hydrogen ion occupies oxygen side. The substitutional hydrogen impurities in $LiNbO_3$ are located between two oxygen atoms of a facet of one oxygen triangle. The hydrogen forms OH bond with the oxygen and it is not co-linear with the O-O bond. Actually, OH bond line slightly deviates from O-O bond line in oxygen plane perpendicular to c-axis and it makes angle $\theta$ (= 10.2°)[22] with the O-O bond in ferroelectric phase. The $H^+$ is bound to one of the $O^{2-}$ neighbors of a lithium vacancy ($V_{Li}$). The impurity hydrogen shows localized vibration which is in the frame work of an anharmonic oscillator model. In anharmonic oscillator model the potential energy will be Morse potential:

$$\phi(r) = D(1 - e^{a(r-r_o)})^2, \qquad (1)$$

where $D$ is the potential depth, $r_o$ is the O-H bond length with lowest potential energy and $a$ be the stretching constant inverse of the length. The substitutional hydrogen is in the vicinity of $i$-th lithium vacancy at the position Q is shown in Fig. 1. The distance AQ is $r$ and $r_o$ be the equilibrium distance denoted by AP. The $i$-th substitutional hydrogen is displaced along AQ line to set up anharmonic oscillation in oxygen plane. At any instant hydrogen displacement is $\eta_i = (r - r_o)$. The displacement of hydrogen is explained by stretching of OH bond in this context. To express the dynamics of $i$-th substitutional hydrogen in this potential, the Hamiltonian of the system is given by

$$H = \frac{p_i^2}{2\mu} + \phi(r), \qquad (2)$$

where $p_i = \mu \dot{\eta}_i$ and $\mu = m_H m_o / (m_H + m_o)$ is the reduced mass of substitutional hydrogen. The symbols $m_o$ and $m_H$ are atomic mass of oxygen and hydrogen respectively. By considering the anharmonic vibrational energy [23] eigen value $E_n$, one can write the following Hamiltonian equation

$$H\psi = E_n \psi \qquad (3)$$

Along OH bond in ferroelectrics LiNbO$_3$, the anharmonicity associated with Morse potential typically yields damped features. The equation of motion of $i$-th substitutional hydrogen with damping is given by

$$\mu \ddot{\eta}_i + \mu \gamma_{OH} \dot{\eta}_i + 2Da(e^{-a\eta_i} - e^{-2a\eta_i}) = 0. \qquad (4)$$

Here $\gamma_{OH}$ is the damping constant. To investigate substitutional diffusion coefficient of hydrogen, we have introduced white noise in the proposed model by including Langevin force in equation (4). Under this condition, the equation (4) may be rewritten as

$$\mu \ddot{\eta}_i + \mu \gamma_{OH} \dot{\eta}_i + 2Da(e^{-a\eta_i} - e^{-2a\eta_i}) = \mu \Gamma(t). \qquad (5)$$

The Langevin force $\mu \Gamma(t)$ is assumed to be a Gaussian random process with δ correlation. The equation (5) is rewritten as

$$\ddot{\eta}_i + \gamma_{OH} \dot{\eta}_i + \frac{2Da}{\mu}(e^{-a\eta_i} - e^{-2a\eta_i}) = \Gamma(t). \qquad (6)$$

To obtain the Fokker Planck equation, the above equation (6) is written as a system of two first order equations mentioned below

$$\dot{\eta}_i = v_i, \qquad (7a)$$

$$\dot{v}_i = -\gamma_{OH} \dot{\eta}_i + \frac{2Da}{\mu}(e^{-a\eta_i} - e^{-2a\eta_i}) + \Gamma(t). \qquad (7b)$$

The substitutional hydrogen moves with velocity $v_i$ along OH bond. The quantity $\Gamma(t)$ is the rapidly fluctuating random term. The average value of Langevin force is zero ($<\Gamma(t)> = 0$) and considering white noise [24] it can be represented as

$$<\Gamma(t)\Gamma(t')> = \frac{2\gamma_{OH} K_B T}{\mu} \delta(t - t'). \qquad (7c)$$

The Fokker Planck equation corresponding to equation (6) is given below

$$\frac{\partial w}{\partial t} = \left[ -\frac{\partial v_i}{\partial \eta_i} + \frac{\partial}{\partial v_i} \left( \gamma_{OH} v_i + \frac{2Da}{\mu} (e^{-a\eta_i} - e^{-2a\eta_i}) \right) + \frac{\gamma_{OH} K_B T}{\mu} \frac{\partial^2}{\partial v_i^2} \right] w. \tag{8}$$

Here $w(\eta_i, v_i, t)$ is the distribution function in position and velocity space, $K_B$ is the Boltzmann constant. The different participating variables of equation (8) along with the spatial term are taken in the non-dimensional form as

$$a\eta_i = \bar{\eta}_i, \tag{9a}$$

$$\frac{\gamma_{OH}}{\omega_o} = \bar{\gamma}_{OH}, \tag{9b}$$

$$\omega_o t = \bar{t}, \tag{9c}$$

$$\frac{v_i}{v_o} = \bar{v}_i, \tag{9d}$$

where $\omega_o = \sqrt{2Da^2/\mu}$ is the characteristic frequency of the vibrational OH bond about $r_o$ in LiNbO$_3$. The impurity hydrogen overcome the Morse potential depth with velocity $v_o = \omega_o/a$. By using equations (9) the equation (8) is reduced to the non-dimensional Fokker Planck equation as given by

$$\frac{\partial w}{\partial \bar{t}} = \left[ -\frac{\partial \bar{v}_i}{\partial \bar{\eta}_i} + \frac{\partial}{\partial \bar{v}_i} \left( \bar{\gamma}_{OH} \bar{v}_i + (e^{-\bar{\eta}_i} - e^{-2\bar{\eta}_i}) \right) + \frac{\bar{\gamma}_{OH} K_B T}{\mu v_o^2} \frac{\partial^2}{\partial \bar{v}_i^2} \right] w. \tag{10}$$

From equation (10), the diffusion coefficient of substitutional hydrogen in LiNbO$_3$ ferroelectrics is written as

$$D_H = \frac{\bar{\gamma}_{OH} K_B T}{\mu v_o^2} = \frac{\gamma_{OH} K_B T a^2}{\mu \omega_o^3}. \tag{11}$$

The equilibrium distribution indicates $\frac{\partial w}{\partial \bar{t}} = 0$. Under this condition the equation (10) becomes time independent and it can be written as

$$\left[ -\frac{\partial \bar{v}_i}{\partial \bar{\eta}_i} + \frac{\partial}{\partial \bar{v}_i} \left( \bar{\gamma}_{OH} \bar{v}_i + (e^{-\bar{\eta}_i} - e^{-2\bar{\eta}_i}) \right) + \frac{\bar{\gamma}_{OH} K_B T}{\mu v_o^2} \frac{\partial^2}{\partial \bar{v}_i^2} \right] w = 0. \tag{12}$$

The solution of equation (12) is given below

$$w = A e^{-\frac{1}{2\chi} \left( \bar{v}_i^2 + (1 - e^{-\bar{\eta}_i})^2 \right)}, \tag{13}$$

where $\chi = K_B T a^2 / \mu \omega_o^2$. To determine the value $A$ the normalization condition is written as

$$A \int_{-\infty}^{\infty} e^{-\frac{1}{2\chi}(\bar{v}_i^2)} d\bar{v}_i \int_{-\infty}^{\infty} e^{-\frac{1}{2\chi}(1-e^{-\bar{\eta}_i})^2} d\bar{\eta}_i = 1. \tag{14}$$

After evaluating the integral $\int_{-\infty}^{\infty} e^{-\frac{1}{2\chi}(\bar{v}_i^2)} d\bar{v}_i$ we get $\sqrt{\frac{2\pi K_B T a^2}{\mu \omega_o^2}}$ and the remaining integral may be written as $\int_{-\infty}^{\infty} e^{-\frac{1}{2\chi}(1-e^{-\bar{\eta}_i})^2} d\bar{\eta}_i = \int_{-\infty}^{\infty} e^{-\frac{1}{2\chi}(1-1+\bar{\eta}_i)^2} d\bar{\eta}_i = \sqrt{\frac{2\pi K_B T a^2}{\mu \omega_o^2}}$. Applying these results in equation (14) we get $A = \frac{\mu \omega_o^2}{2\pi K_B T a^2}$. So equation (13) becomes

$$w = \frac{\mu \omega_o^2}{2\pi K_B T a^2} e^{-\frac{1}{2\chi}\left( \bar{v}_i^2 + (1-e^{-\bar{\eta}_i})^2 \right)}. \tag{15}$$

With the use of equations (2), (8), (13) the equation (15) can be written as

$$w = \frac{\mu \omega_o^2}{2\pi K_B T a^2} e^{-\frac{E_n}{K_B T}}. \tag{16}$$

This is the Boltzmann distribution equation. Where the vibrational energy levels are written as

$$E_n = \hbar \omega_o \left( n + \frac{1}{2} \right) - \frac{\hbar^2 \omega_o^2}{4D} \left( n + \frac{1}{2} \right)^2. \tag{17}$$

The substitutional hydrogen makes OH bond that is initially considered to be in ground state having energy $E_o$ for $n = 0$. The percentage of substitutional hydrogen atoms are that of OH bonds which are calculated by multiplying 100 with the probability $w$ given in equation (16). The Percentage of formation of OH bond in ferroelectrics are strongly depended on temperature. The characteristic frequency for all the OH bonds in ferroelectrics are not identical. To illustrate the Fokker Planck equation (10) we introduced number of OH bonds at different temperature for finite time is expressed as

$$\rho = e^{-\left(\frac{\frac{\bar{v}_i^2}{2}+V(\bar{\eta}_i)}{\chi}\right)} e^{-\bar{\gamma}_{OH}\bar{t}}, \tag{18}$$

where $V(\bar{\eta}_i) = \frac{1}{2}(1 - e^{-\bar{\eta}_i})^2$. The distribution function $w(\bar{\eta}_i, \bar{v}_i, \bar{t})$ can be expressed as

$$w(\bar{\eta}_i, \bar{v}_i, \bar{t}) = U(\rho, \bar{t}) e^{\bar{\gamma}_{OH}\bar{t}}, \tag{19}$$

$$\frac{\partial W}{\partial \bar{t}} = \frac{\partial W}{\partial \bar{t}} + \frac{\partial W}{\partial \rho}\dot{\rho}. \tag{20}$$

With the use of equations (18), (19), (20) the non-dimensional Fokker Planck equation (10) is expressed through coordinate transformation in terms of $\rho$ and $\bar{t}$ as

$$\frac{\partial U}{\partial \bar{t}} = \bar{\gamma}_{OH} \frac{\bar{v}_i^2}{\chi} \frac{\partial^2 U}{\partial \rho^2} e^{-2\left(\frac{\frac{\bar{v}_i^2}{2}+V(\bar{\eta}_i)}{\chi}\right)} e^{-2\bar{\gamma}_{OH}\bar{t}}. \tag{21}$$

Now we introduce the new parameter $\theta$ by the following expression

$$\theta = \frac{\bar{v}_i^2}{2\chi} e^{-2\left(\frac{\frac{\bar{v}_i^2}{2}+V(\bar{\eta}_i)}{\chi}\right)} (1 - e^{-2\bar{\gamma}_{OH}\bar{t}}). \tag{22}$$

With this new parameter $\theta$ the non-dimensional Fokker Planck equation (21) is reduced to Fick's equation as

$$\frac{\partial U}{\partial \theta} = \frac{\partial^2 U}{\partial \rho^2}. \tag{23}$$

The solution of this equation is given by

$$U(\rho, \bar{t}) = \frac{1}{\sqrt{4\pi\theta}} e^{-\rho^2/4\theta}. \tag{24}$$

Using $\rho$ and $\theta$ from equations (18) and (22) the solution can be represented as

$$U = \frac{1}{\sqrt{4\pi(1-e^{-2\bar{\gamma}_{OH}\bar{t}})} \frac{\bar{v}_i}{\sqrt{2\chi}} \exp\left(-\frac{\frac{\bar{v}_i^2}{2}+V(\bar{\eta}_i)}{\chi}\right)} \exp\left(-\frac{e^{-2\bar{\gamma}_{OH}\bar{t}}}{\frac{2\bar{v}_i^2}{\chi}(1-e^{-2\bar{\gamma}_{OH}\bar{t}})}\right). \tag{25}$$

Next, dropping the non-dimensional parameters the equation (25) is written as equation (26) to represent the particle density as a function of time and velocity.

$$U = \frac{1}{\sqrt{4\pi(1-e^{-2\gamma_{OH}t})} \sqrt{\frac{\mu v^2}{2 K_B T}} \exp\left(-\frac{E_n}{K_B T}\right)} \exp\left(-\frac{e^{-2\gamma_{OH}t}}{\frac{2\mu v^2}{K_B T}(1-e^{-2\gamma_{OH}t})}\right). \tag{26}$$

The substitutional hydrogen atoms are participated in the process of the formation of OH bonds. The formation of OH bonds are distributed in such way that strongly depends on velocity of substitutional hydrogen and time as indicated in equation (26).

## 3. Results and discussion

The diffused hydrogen in LiNbO$_3$ produced two types of impurity defects: substitutional and interstitial. The substitutional hydrogen which substitutes lithium and forms OH bond with the oxygen of a facet of oxygen-triangle neighbors of a lithium vacancy in oxygen plane perpendicular to c-axis. On the other hand, the interstitial hydrogen which is commonly found in metal and semiconductor. The diffusion coefficient of interstitial hydrogen is greater than substitutional hydrogen since the size of the interstitial hydrogen is smaller than substitutional hydrogen [26]. The diffusion coefficient of interstitial hydrogen is greater than that of substitutional hydrogen. The OH bond line makes an angle 10.2° with the O-O bond line. The equilibrium OH bond length is $r_o = 0.988 Å$ [28]. The stretching mode of OH bond is speculated the anharmonic oscillation. To explain this phenomenon the Morse potential is adopted in the proposed model. The Morse potential according to equation (1) is shown in Fig. 2 with the potential depth, $D = 4.43\ eV\ (7.088 \times 10^{-12} ergs)$ and stretching constant, $a = 2.283 \times 10^8$ cm$^{-1}$. The reduced mass of hydrogen is $\mu = 1.573 \times 10^{-24}$ gm. The classical vibrational frequency of OH bond due to substitutional hydrogen in oxygen plane is $\omega_o = 6.88 \times 10^{14}$ Hz. The average value of Langevin force is zero as it is random in nature. Introducing white noise in Fokker Planck equation is employed to explain the thermal fluctuation. The expression of diffusion coefficient is derived from the non-dimensional Fokker Planck equation for LiNbO$_3$. The experimentally reported value of diffusion coefficient is $2.1 \times 10^{-12} cm^2/s$ [25] at 360°C and $4.6 \times 10^{-12} cm^2/s$ [26] at 400°C. Applying the above data, the relaxation time is calculated from equation (11) as $\tau = 2.07\ \mu s$ which is the reciprocal of damping constant.

The concentration profile of hydrogen due to substitution of lithium before annealing is about 90% or more. Upon annealing for 1hour this level drops to nearly 50% [25]. The surface concentration and diffusion coefficient of interstitial hydrogen and substitutional hydrogen in proton exchange LiNbO$_3$ layer annealed at 400°C are nicely reported by Casey et al. [26] and by applying their data one can calculate the percentage of substitutional hydrogen which lies in the range 73.68% to 69.69% corresponding to the annealed time 6 min to 180 min. Another experimental study shows that the structural disorders are introduced in LiNbO$_3$ due to proton exchange where roughly 75% of lithium cations leave the layer and are exchanged by proton [27]. The distribution of particles has been derived from non-dimensional Fokker Planck equation under equilibrium condition in equation (16). The OH bonds initially are considered in ground state having energy $E_o = 0.224\ eV$. According to the proposed model equation (16) shows that the percentage of substitutional hydrogen is 52.35% at 400°C and 74 % at 500°C. The variation of substitutional hydrogen is shown in Fig. 3 with temperature. This analytical results agree with the experimental results reported in refs. [26] and [27]. The number of substitutional hydrogens (OH bonds) depended on the velocity and time as shown in Fig. 4 according to the equation (26). The damping constant of OH bond is investigate theoretically to calculate the relaxation time. The damping constant has an important role in ferroelectrics to determine the refractive index due to impurity and other properties of ferroelectrics.

## 4. Conclusions

In the present article we have studied analytically the diffusion coefficient, relaxation time and percentage of substitutional hydrogen in LiNbO$_3$ from damped anharmonic oscillation model by using Morse potential. The variation of substitutional hydrogen in LiNbO$_3$ with temperature is agreed with the reported experimental values. The variation of substitutional hydrogen with velocity and time is derived from Fokker Planck equation and it has been presented graphically.

**References**

1. M. E. Lines and A. M. Glass, Principles and Applications of Ferroelectrics and Related Materials, Clarendon, Oxford, (1977).
2. S. Kim, V. Gopalan and A. Gruverman, Appl. Phys. Lett. **80**, 2740 (2002)
3. K. T. Gahagan et al., Appl. Opt. **38**, 1186 (1999).
4. S. Kim, V. Gopalan, K. Kitamura and Y. Furukawa, J.Appl. Phys. **90**, 2949 (2001)
5. M. Dawber, K. M. Rabe, and J. F. Scott, Rev. Mod. Phys. **77**, 1083 (2005)
6. A. M. Prokhorov and Yu. S. Kuz'minov, Physics and Chemistry of Crystalline Lithium Niobate, Adam Hilger, Bristol and New York (1990).
7. Y. N. Korkishko and V. A. Fedorov, J. Appl. Phys, **82**, 1010, (1997).
8. R. Hunsperger, Integrated Optics: Theory and Technology, Spinger, New York, (2009).
9. R. G. Smith, D. B. Fraser, R. T. Denton, and T. C. Rich, J.Appl. Phys, **39**, 4600, (1968).
10. J. M. Zavada, H. C. Casey, C. H. Chem, and A. Lono, Appl. Phys. Lett. **62**, 2769 (1993).
11. I. Nee, K. Buse, F. Havenmer, R. A. Rupp, M. Fally, and R. P. May, Phys. Rev. B. **60**, R9896 (1999).
12. L. Ren, L. Liu, D. Liu, J. Zu, Z. Luan, Optical Society of American Journal B **20**, 2162-2173 (2003).
13. L. Kovacs, M. Wohlecke, A. Jovanovic, K. Polgar, and S. Kapphan, J. Phys. Chem. Solids **52** 797 (1991).
14. C. E. Rice, J. Solid State Chem **64**, 188 (1986).
15. T. Volk and M. Wohlecke, Lithium Niobate, Spinger, (2008).
16. V. Bermudez, L. Huang, D. Hui, S. Field, E. Dieguez, Appl. Phys. A **70**, 591 (2000).
17. V. Gopalan, T. E. Mitchell, K. E. Sicakfus, Solid State Commun. **109**, 111 (1999).
18. A. K. Bandyopadhyay, P. C. Roy and V. Gopalan, J. Phys: Condens Matter **18**, 4093 (2006).
19. P. Giri, K. Choudhary, A. Dey, A. Biswas, A. Ghosal and A. K. Bandyopadhyay, Phys. Rev. B **86**, 184101(2012).
20. P. Giri, and M. K. Mandal, AIP Advances **4**, 107140 (2014)
21. W. B Yan, et. al., Phys. Stat. Solidi **201**, 2013 (2004).
22. H. H. Nahm and C. H. Park, Appl. Phys. Lett. **78**, 24 (2001).
23. P. M. Morse, Phys. Rev. **34** (1929)
24. H. Risken, The Fokker Planck Equation Methods of Solution and Applications, Springer 1996
25. G. R. Paz-Pujait, D. D. Tuschel, S. T. Lee, and L. M. Salter, J. Appl. Phys **76**, 3981 (1994)
26. H. C. Casey, C. Chen, J. M. Zavada, S. W. Novak, Appl. Phys. Lett. **63**, 718 (1993).
27. A. Grone and S. Kapphan, J. Phys. Chem Solids **56**, 687 (1994)
28. F. Freytag, G. Corradi and M. Imlau, Scientific Reports **6**, 36929 (2016)

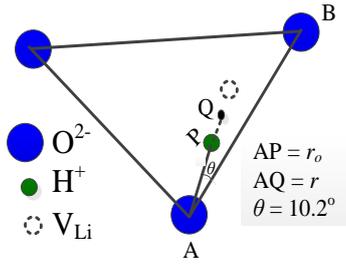
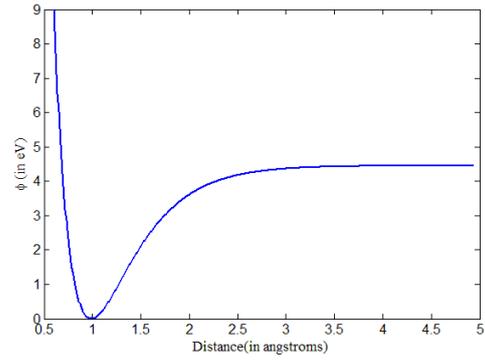

**Fig. 1.** Schematic representation of oxygen plane in LiNbO$_3$ [28].

**Fig. 2.** Morse potential

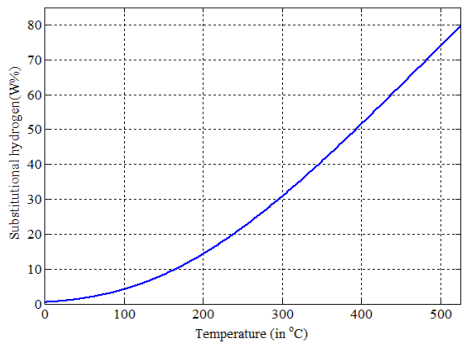
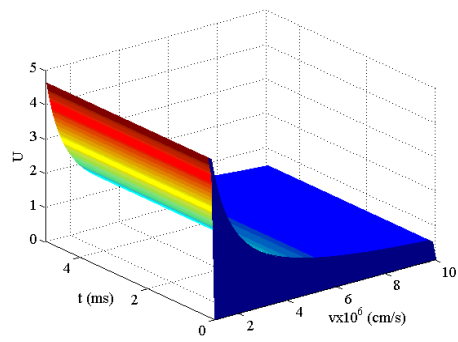

**Fig. 3.** The variation of substitutional hydrogen with temperature in LiNbO$_3$ (Using equation 16).

**Fig. 4.** The variation of U with time and velocity according to equation 26.